# Optical control of waves in a cardiac excitable medium

by

Rebecca A. B. Burton[1], Aleksandra Klimas[2], Christina M. Ambrosi[2],
Emilia Entcheva[2] and Gil Bub[1, #]

[1]Department of Physiology, Anatomy and Genetics, University of Oxford, United Kingdom
[2]Department of Biomedical Engineering, Stony Brook University, Stony Brook, NY, USA

# Corresponding author:
Gil Bub, PhD
Department of Physiology, Anatomy and Genetics
University of Oxford
Sherrington Building, Parks Road, Oxford OX1 3PT
United Kingdom
Email: gil.bub@dpag.ox.ac.uk

**In nature, excitable reaction-diffusion systems found in diverse settings (e.g. chemical reactions, metal rust, yeast, amoeba, heart, brain) generate geometrically similar macroscopic waves[1,2]. For the heart and brain, the spatiotemporal patterns formed by these excitation waves separate healthy from diseased states[1-3]. Current electrical and pharmacological therapies for bioelectric disorders often lack the necessary spatiotemporal precision needed to control these patterns. Optical methodologies have the potential to overcome these limitations, but have only been demonstrated in simple systems, e.g. the Belousov-Zhabotinsky (BZ) chemical reaction[4]. Here we combine novel dye-free optical imaging with optogenetic actuation for dynamic control of cardiac excitation waves. We leverage patterned light to optically control emergent macroscopic properties of cardiac tissue: wave direction, wave speed, and spiral wave chirality. This all-optical approach offers a fundamentally new experimental platform for the study and control of pattern formation in complex biological excitable systems.**

Heart cells form a dense, well-coupled excitable medium (a syncytium), which supports macroscopic propagating waves of activity with characteristic space scales that are orders of magnitude larger than the cells themselves. Wave dynamics and the resultant spatiotemporal patterns are central to heart's function. While planar waves, emanating from a central pacemaking source, act to synchronize contraction during normal heart beat, aberrant re-entrant waves, with characteristic spiral morphology, keep rapidly re-exciting the tissue and underlie potentially deadly tachycardias and fibrillation. Visualizing, understanding and ultimately controlling excitation waves in the heart can aid in providing life-saving solutions.

Optical mapping with fluorescent voltage-sensitive dyes[3] has helped confirm the existence of planar and spiral waves of excitation in cardiac tissue. Beyond observation however, a comparable means for *manipulation* of excitation waves is lacking. From a theoretical point of view, an optical approach using dynamic light patterns to perturb excitation is most likely to achieve the robustness and high spatiotemporal precision needed for fine control. Such active perturbation of excitation waves by light has been demonstrated in photosensitive versions of the BZ chemical reaction[4] and to some degree in light-sensitive Dictyostelium[5] amoeba colonies. However, existing tools for manipulation of excitation waves in living mammalian tissue are very limited: in the heart, waves can be initiated and terminated crudely by electrical or pharmacological means that lack spatial and temporal precision.

Optogenetics, the inscription of light sensitivity in mammalian tissues through the genetic expression of microbial opsins[6,7], holds the promise to enable such fine spatiotemporal targeting of excitation waves. The ability to optically address specific cells and cell types has been leveraged in neurosciences to track and control neural circuits[8], but its use in the cardiac field is only in its infancy[9], with early reports showing its utility in optical initiation[10-12] and termination of cardiac excitation waves[13,14]. Fine control of conduction will benefit from the combined power of optical imaging and spectrally-compatible high-resolution optogenetic perturbation techniques[9,15]. However, to date, optogenetics has exclusively focused on perturbing cell-level properties, while fine control of macroscopic excitation waves has not been demonstrated in neural or in cardiac preparations.

In this letter, we report the combined use of dye-free optical imaging of excitation waves with optogenetic perturbation to control wave propagation and pattern formation in a biological excitable medium. Cardiac monolayers of coupled primary cardiomyocytes, known to support classic travelling excitation waves[16], are genetically modified to uniformly express channelrhodopsins, adding optical responsiveness without otherwise altering innate functionality[17]. Patterned blue light is used for the precise optical manipulation of key wave properties: conduction velocity, wave direction, and spiral wave chirality, not otherwise achievable by any conventional methods.

We use a spectrally-flexible, dye-free optical imaging modality that is easily combinable with spatiotemporal optogenetic perturbations. The technique relies on cellular excitation-contraction induced changes in the optical path length (OPL) along the sample's z-axis (normal) to generate time-dependent interference images. We use oblique trans-illumination from a partially-coherent LED light source (**Fig 1a**) to perform fast imaging (using a sCMOS camera) of OPL changes allowing for visualization of activation wavefronts at the macro-scale directly or with minimal post-processing, without dyes and special optics (**Fig 1b**). In addition to the rejection of direct rays (0$^{th}$ order), some degree of coherence (LED or a laser) is essential for viewing microscopic cellular and subcellular OPL changes over a macroscopic field of view of 1 to 4cm$^2$, which suggests that the mechanism of image formation includes interference of the illumination beam with light dynamically deviated by cell structures. Additionally, the coherence length of LEDs used in this study is comparable to the axial thickness of the sample (~10μm), suggesting the cells act as local interferometers sensitive to time-varying OPL changes. Furthermore, activation wave dynamics remain visible when the sample is defocused, consistent with the generation of a holographic image outside the sample plane[18-20]. Interestingly, an alternative dye-free imaging configuration was reported by Lee and colleagues [21] ten years ago using incoherent light, spatially filtered by a pinhole, to visualize macroscopic wave dynamics in cardiac monolayers. Further work is required to determine and compare the exact mechanism of image formation of these two systems. Key advantages of the dye-free imaging reported here, in contrast to fluorescence-based approaches[9,15], include flexibility in wavelength choice when combined with optogenetic actuation, the use of low-light levels, simple (low-NA) optics and conventional cameras. These features enable the non-invasive visualization of wave propagation during optogenetic actuation (**Fig 1a**), realized here by the delivery of dynamic spatiotemporal patterns using a blue LED projected through a computer-controlled digital micromirror device (DMD). The utility of the developed all-optical system is illustrated by three proof-of-concept examples of optical control of wave properties.

In homogeneous cardiac tissue, a stimulus triggers uniform wave propagation in all directions. However, certain conditions can temporarily block the wave in a particular direction. Such unidirectional block enables the wave to curl and circle back on itself, leading to re-entry and lethal cardiac arrhythmias[22,23]. Experimentally, unidirectional block can be recreated by generating a wavefront that interacts with the wake of a pre-existing wave, but reliable results are difficult to obtain with conventional electrical stimulation. Here, we use light to generate unidirectional block of wave propagation in the genetically modified cardiac syncytium without the need of interaction with a pre-existing wavefront. Dosed optogenetic stimulation can "pre-clamp" the affected cells to a depolarized (non-excitable) state; timed release from this state marks the start of a refractory period, after which the cells are excitable again. **Fig 2** illustrates that timed, space-defined removal of light, after a global depolarizing optical clamp, results in control of tissue refractoriness and selection of the exact time and location of unidirectional block upon new stimulation: bidirectional propagation is triggered in **Fig 2b**, and right or left unidirectional block are demonstrated in **Fig 2c-d**, using asymmetric light release.

Wave conduction velocity plays a key role in excitable media dynamics as it directly influences the spatial scale that can accommodate a re-entrant circuit[24]. Experimentally, only crude pharmacological tools exist to influence cardiac conduction velocity (largely defined by the cell-cell coupling)[25]. Here we report that low-light application can speed up propagation - dosed subthreshold depolarization brings cells closer to the threshold for excitation, and yields shorter activation times. Conduction velocity can be increased in user-defined regions of the tissue by applying low light ahead of a triggered wave (**Fig 3**). Conduction velocity is increased first to the left (**Fig 3b**), and then to the right (**Fig 3c**) without affecting propagation properties in unilluminated regions of the tissue. We were able to precisely control conduction velocity by varying the light levels – linear velocity increase is shown over a wide range of light intensities (**Fig 3d**).

Finally, spiral waves are a prototypical example of self-organization in distributed excitable media[1,2], including autocatalytic chemical reactions, yeast, amoeba colonies, heart, cortical and retinal preparations. Experimental control of spiral wave dynamics is an important tool for understanding emergent pattern formation in excitable media, and, in living systems, it may lead to clinically-relevant therapeutic solutions. Effective control strategies require the ability to deliver precise spatially-distributed perturbations. This presents a challenge for controlling cardiac spiral waves by conventional electrical means lacking such capabilities. Recently, global optogenetic stimulation was used to abolish cardiac spiral waves[12,14]. Here, we demonstrate a finer level of optical control, exemplified by light-controlled reversal of cardiac spiral wave chirality. The spiral chirality (direction of rotation) is a fundamental property that affects how the wave interacts with other wavefronts as well as the underlying medium[26,27]. To our knowledge, controlled modulation of spiral wave chirality has not been demonstrated experimentally in any excitable system, though theoretical ideas to do so have been proposed[28]. To reverse the chirality of an ongoing spiral wave, we transiently impose a computer-generated (by a cellular automaton model) counter-rotating spiral wave of shaped light with slightly higher frequency of rotation compared to the native spiral (**Fig 4a**). Within 1-2 rotations, the imposed wave effectively overwrites the existing spiral. The shaped refractory gradient left after light removal perpetuates the spiral by allowing the light-triggered wave to persist and re-enter along the path of the imposed spiral. Interestingly, the spirals drift over a few rotations to a preferred location that is different for clockwise and counter-clockwise spirals, and the latter are faster (18%), likely indicating asymmetry in the underlying tissue microstructures. The chirality control was robustly deployed repeatedly resulting in the same outcome, independent of the phase of the ongoing spiral (**Fig 4b,c**).

In summary, we present a new all-optical framework for modulation of cardiac excitation waves, in ways that are impossible by conventional techniques. We demonstrate the power of optogenetics to control wave pattern formation in cardiac tissue at a level normally associated with computer models or simpler experimental systems - photosenstive chemical reactions.  Furthermore, since diverse excitable media display similar macroscopic dynamics, insights from the optogenetically modified cardiac syncytia described here will be relevant to a broad range of biological and chemical reaction-diffusion systems. The ability to precisely control light will enable new research on pattern formation in complex biological excitable media.

**Acknowledgements**
We thank Dr Harold Bien, Dr Alex Corbett, Mr Jakub Tomek and Mr Suhail Aslam for helpful discussions and technical assistance. GB acknowledges support from the BHF Centre of Research Excellence, Oxford (RE/08/004). R.A.B.B. is funded by an EPSRC Developing Leaders Grant, a Goodger award and holds a Winston Churchill Fellowship and Paul Nurse Junior Research Fellowship (Linacre College, Oxford). This work was supported by MR/K015877/1 (G.B.), NIH R01 HL111649 (E.E.), and a NYSTEM grant C026716 to the Stony Brook Stem Cell Centre.

**Author Contributions**

G.B. and E.E. initiated the project and provided guidance. R.A.B.B. and G.B. performed the experiments.  G.B. wrote the software to collect and analyse the data. C.M.A. and E.E. developed and provided biological materials and guidance on the optogenetic manipulations. R.A.B.B. and A.K. helped with data interpretation and figure preparation. G.B. and E.E. wrote the manuscript. All authors were involved in analysis of the results and revision of the manuscript.

**FIGURES**

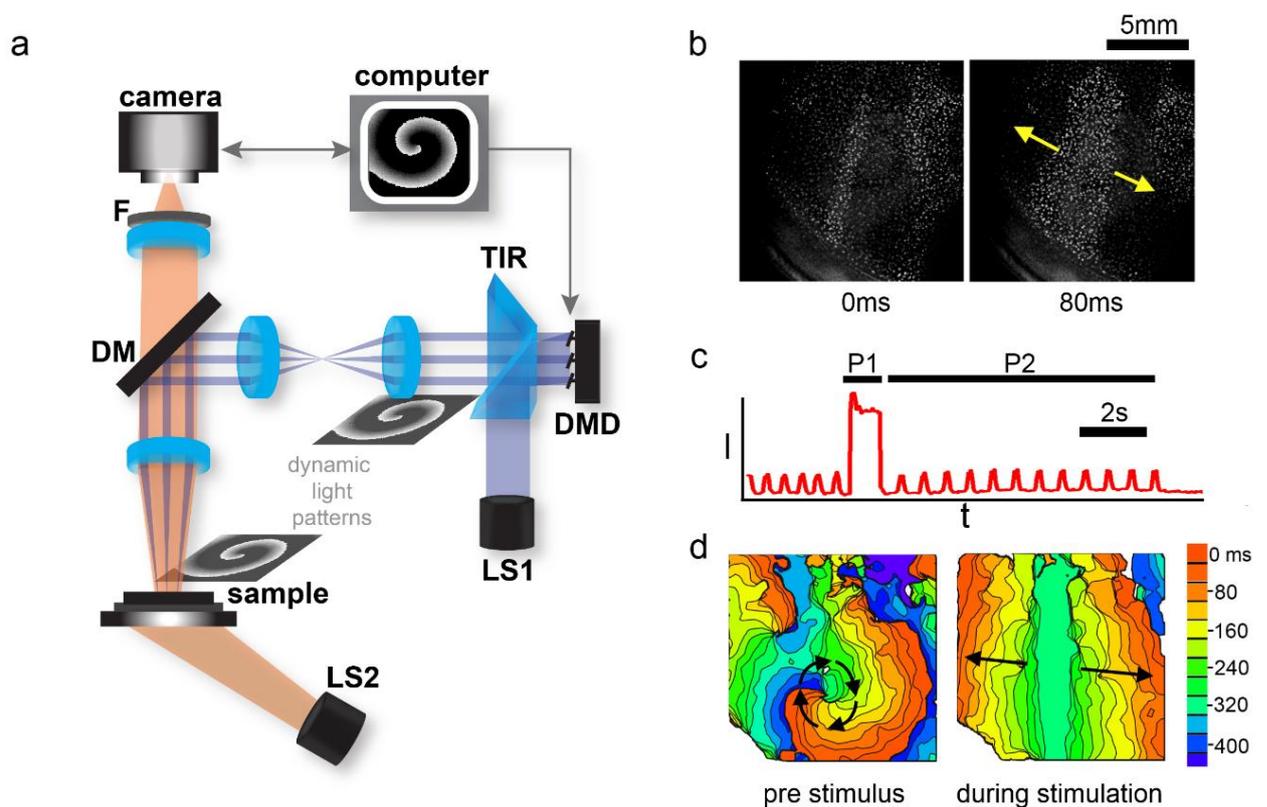

**Figure 1 All-optical system for space-time resolved control of wave dynamics in biological media:**
**a.** Experimental setup, including an actuation light source, LS1 (10W LED, 460nm), a total-internal-reflection prism (TIR) and a computer-controlled digital micromirror device (DMD). Generated dynamic light patterns are projected via lenses and a dichroic mirror, DM (510nm) to the biological sample. A second light source, LS2 (white LED, with bandpass filter at 580±20 nm) provides oblique trans-illumination for dye-free imaging onto a sCMOS camera through a low-NA objective lens (1x, 0.25 NA) and a long pass emission filter, F (>580nm). **b.** Example of minimally filtered images in response to optical line stimulation in cardiac monolayers; **c.** and **d.** Temporal trace from a single pixel and activation maps showing ongoing spontaneous activity (a spiral) pre-stimulus, terminated by a strong global optical stimulation (P1), and followed by P2 periodic optical stimulation by a line stimulus.

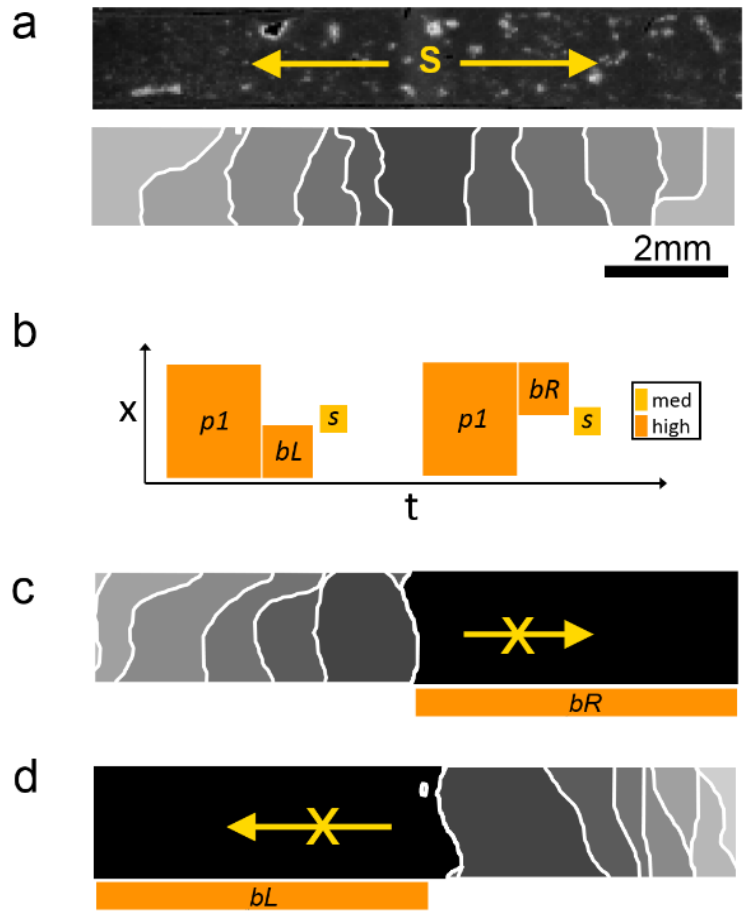

**Figure 2 Optical control of wave direction in cardiac monolayer: a.** sample and applied light stimulus S, inducing bi-directional propagation; **b.** schematic representation of the applied light protocol in space-time (x-t) with pre-conditioning stimuli p1, and blocking stimuli bL and bR to set tissue refractoriness prior to stimulus S, resulting in a right-side or a left-side unidirectional block (**c** and **d**). Here p1 is 350 ms, bR and bL are 50 ms, and S is 10 ms. Activation maps show isochrones at 100 ms spacing.

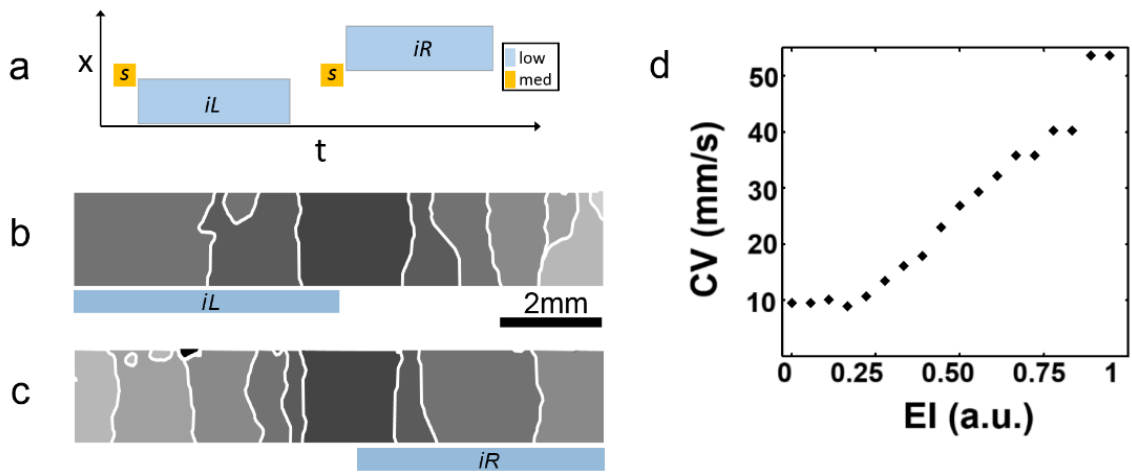

**Figure 3 Optical control of cardiac wave conduction velocity: a.** schematic of applied optical stimulation protocol in space (x) and time (t); **b-c**. activation maps of controlled left-side (**b**) and right-side increase of conduction velocity (**c**) by light, indicated by the larger spacing of the isochrones; **d.** relationship between conduction velocity (CV) and relative irradiance magnitude (EI). Linear relationship was confirmed in n=5 samples. iR and iL are 500 ms. Isochrones are 100 ms apart.

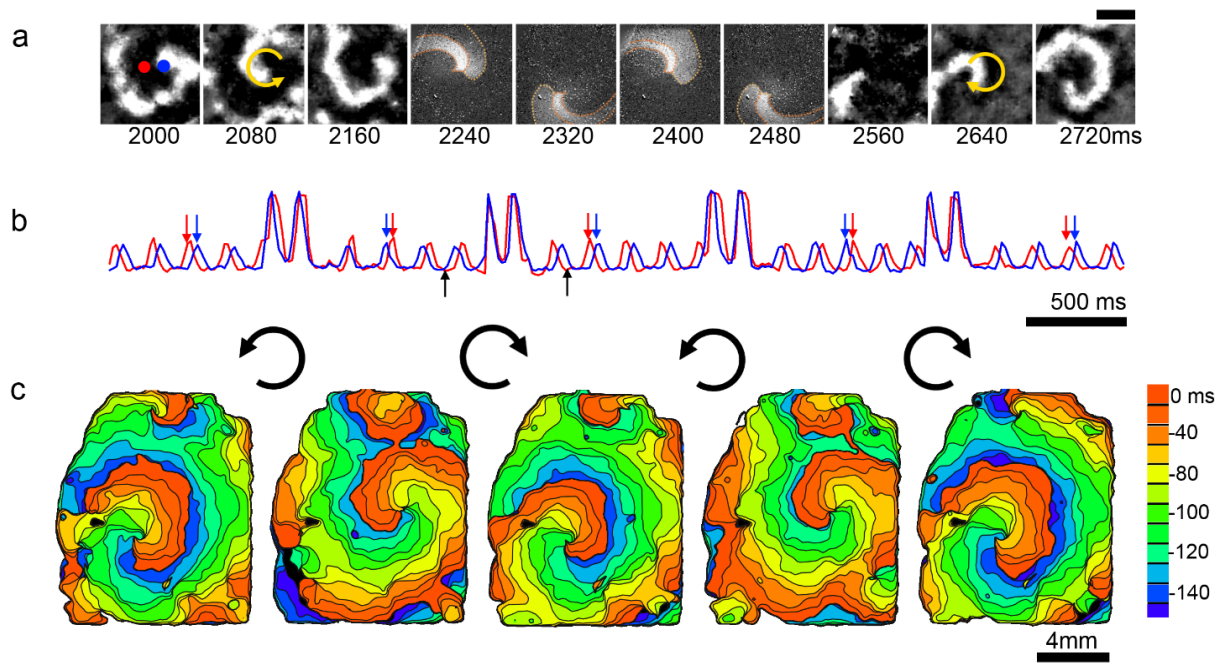

**Figure 4 Optical control of spiral wave chirality in cardiac syncytium: a.** Snapshots from an ongoing counter-clockwise spiral wave (frames 2000-2160), an optically applied computer-generated clockwise spiral wave (frames 2240-2480) and the persisting spiral wave post-chirality reversal (frames 2560-2720). **b.** Temporal traces of activity from the indicated red and blue pixels showing four light-controlled chirality reversals (computer-generated spirals were faster and imposed at random phase for less than two rotations, as seen in the four higher-intensity transients; black arrows indicate the time period presented in panel (**a**); red and blue arrows indicate the switch of order of excitation at the chosen locations due to chirality reversal. **c**. activation maps for the initial spiral wave and the four resultant spirals after each of the chirality reversals by light. Note slight difference in spiral wave tip location and in rotation frequency for clockwise vs. counter-clockwise spirals.